\documentclass[sort&compress,aps,prd,5p,superscriptaddress,showpacs,showkeys]{elsarticle}
\usepackage{amsmath,amssymb,color,graphicx,bm,braket,booktabs}
\usepackage[toc,page]{appendix}

\newcommand{\1}[1]{\, \mathrm{#1}}
\newcommand{\n}[1]{\mathrm{#1}}

\begin{document}
\title{Improved Pulse Shape Discrimination in EJ-301 Liquid Scintillators}

\author[purdue]{R.F.~Lang}\ead{rafael@purdue.edu}
\author[purdue]{D.~Masson}\ead{dmasson@purdue.edu}
\author[purdue]{J.~Pienaar}\ead{jpienaa@purdue.edu}
\author[ptb]{S.~R\"ottger}\ead{stefan.roettger@ptb.de}
\address[purdue]{Department of Physics and Astronomy, Purdue University, West Lafayette, USA}
\address[ptb]{Physikalisch-Technische Bundesanstalt, Braunschweig, Germany}

\begin{abstract}
Digital pulse shape discrimination has become readily available to distinguish nuclear recoil and electronic recoil events in scintillation detectors. We evaluate digital implementations of pulse shape discrimination algorithms discussed in the literature, namely the Charge Comparison Method, Pulse-Gradient Analysis, Fourier Series and Standard Event Fitting. In addition, we present a novel algorithm based on a Laplace Transform. Instead of comparing the performance of these algorithms based on a single Figure of Merit, we evaluate them as a function of recoil energy. Specifically, using commercial EJ-301 liquid sctintillators, we examined both the resulting acceptance of nuclear recoils at a given rejection level of electronic recoils, as well as the purity of the selected nuclear recoil event samples. We find that both a Standard Event fit and a Laplace Transform can be used to significantly improve the discrimination capabilities over the whole considered energy range of $0-800 \1{keV_{ee}}$. Furthermore, we show that the Charge Comparison Method performs poorly in accurately identifying nuclear recoils.
\end{abstract}

\maketitle
%\linenumbers
\section{Introduction}

\begin{figure*}[htbp!]
  \includegraphics[width=\textwidth,height=4cm]{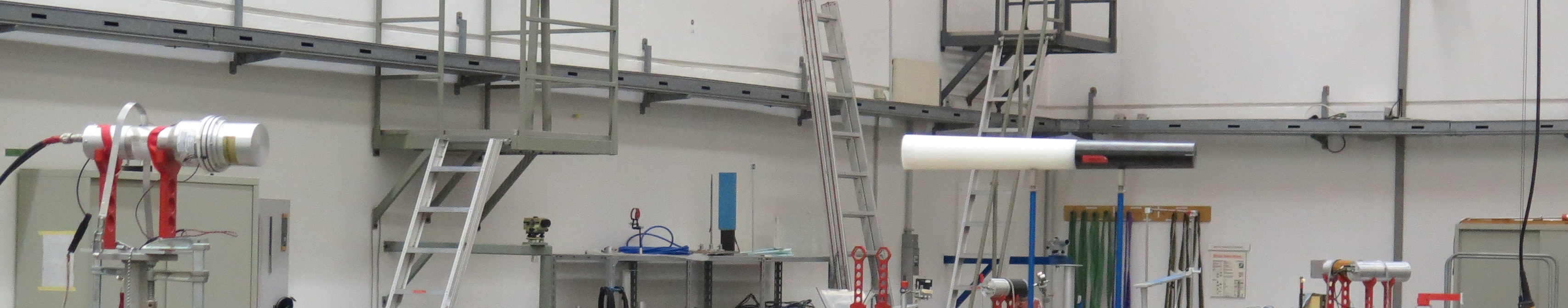}
  \caption{The irradiation setup of the EJ-301 detector (left) corresponding to the 0$^{\circ}$ orientation in Table \ref{table:data_sets}. The shadow cone is visible toward the right. The neutron beam enters the setup from the right.}
\end{figure*}\label{fig:detector_setup}

Liquid scintillators such as EJ-301 (which is similar to NE-213 and BC-501) are very popular for neutron detection as they can easily be shaped into the desired size and geometry of a given application and offer fast timing performance. However, since such liquid scintillators are also sensitive to gamma rays, pulse-shape discrimination (PSD) techniques are essential in order to correctly identify neutron interactions in the detector.

The ability to discriminate nuclear recoil (NR) events from electronic recoil (ER) events originates in the particular production mechanisms of scintillation light in organic liquid scintillators. These liquids are aromatic compounds which have planar molecular structures built up from benzenoid rings. Such structures allow for extended groupings of conjugated molecular bonds between unsaturated carbon atoms~\cite{Brooks1979}. This results in some of the valence electrons of the carbon atoms being delocalized in $\pi$-molecular orbitals. It is the excitations of these $\pi$-electronic states that create the fluorescence observed in organic scintillators. During these excitations, $\pi$-electrons can be promoted from the ground state~S$_0$ to excited singlet states~S$_n$ or triplet~T$_n$ states. For low excitation densities, all excited singlet states above the first excited singlet states S$_1$ decay rapidly and non-radiatively to the lowest excited singlet state. This state then decays exponentially producing fluorescence in the process.

In contrast, the decay of the triplet state is governed by the diffusion time-scale of the triplet exciton and results in delayed fluorescence in which the intensity does not decay exponentially. NRs exhibit greater energy-loss rates and thus have higher densities of triplet states. Pulses from the ionization tracks of these particles exhibit higher yields of delayed fluorescence, hence decaying more slowly than those of ERs. Scintillation light from EJ-301 has three main decay components: $3.2\1{ns}$, $32\1{ns}$ and $270\1{ns}$~\cite{Kuchnir1968}. The slowest of these decay times is produced by the delayed fluorescence of triplet states.

The different pulse shapes that arise from electronic and nuclear recoils in liquid scintillators can be exploited using different PSD techniques. The most popular techniques applied are the Charge Comparison Method~\cite{Brooks1959} and the Zero Crossing Method~\cite{Alexander1961}. These methods were originally implemented in purpose-designed analogue electronics~\cite{Adams1978}, but with the advent of greater computing power at reduced costs, these techniques have been implemented digitally~\cite{Kaschuck2005, Cester2014, Liao2014}. Digital capture of the full waveform allows for offline processing of events, reducing dead time in data acquisition systems. Techniques designed for analogue circuits do not take advantage of the increased information available in the digital domain. Consequently, new PSD techniques have been developed recently~\cite{DMellow2007, Gamage2011, Yousefi2009}. These techniques offer new PSD approaches in the time domain of the waveform, allow frequency-domain and decay-time differences to be investigated using wavelet analysis, and can implement Fourier and Laplace transforms.

Traditionally, the performance of PSD techniques is characterized using the Figure of Merit (FOM), defined as:
\begin{equation}\label{eq:FOM}
\text{FOM} = \frac{\text{Peak Seperation}}{\text{FWHM}_\gamma+\text{FWHM}_n}
\end{equation}
where peak separation refers to the distance between the center of the neutron and gamma distributions in a histogram of the discrimination parameter, and $\text{FWHM}_i$ is the full-width half maximum of the respective distributions. Hence, the FOM does not provide any information on the energy dependence of the performance of PSD techniques. This precludes a comparison of the various algorithms across different authors that may use different energy thresholds in the calculation of their FOM, and additionally, may mask performance issues of the algorithms in particular at low recoil energy. Therefore, we examined the energy-dependent ability of PSD techniques to discriminate between ER and NR events. Furthermore, we determined the efficiency of EJ-301 for detection of neutrons as a function of energy.

\section{Setup}

The fast neutron detector used in this work is a 3"~cell of EJ-301 liquid organic scintillator optically coupled to a fast photomultiplier tube (PMT), type 9821KB manufactured by ET Enterprises. The detector response to neutrons was characterized at the Physikalisch-Technische Bundesanstalt (PTB) in Braunschweig, Germany,  using a deuterium ion beam hitting a Ti($^3$H) target. The deuterium ion beam energy (3.356 MeV) was chosen to produce $(2.500\pm 0.010)\1{MeV}$ (k=1 according to~\cite{GUM}) monoenergetic neutrons via $^{3}$H(d,n)$^{4}$He nuclear reaction, in the direction of the ion beam. The detector was placed $3\1{m}$ from the target. The output of the PMT was connected to a CAEN DT5751 digitizer, which samples at 1~GHz with a  resolution of 10~bits. This digitizer has a $1\1{V}$ dynamic range. A $1\1{MeV_{ee}}$ pulse from an ER event in the PMT produces a $550\1{mV}$ signal.

Data were collected at three different nominal beam current settings to study the effect of neutron flux on the performance of the detector. The detector was placed such that the neutron beam was parallel to the normal of the front face, defined as an angle of 0$^{\circ}$. The distance between the front face of the detector and the active layer of the target was $(3000\pm 2) \1{mm}$ (k=2~\cite{GUM}) for all measurements. Additional data were taken at each setting with a shadow cone, made of iron and polyethylene, placed between the target and the detector to measure the in-scatter of neutrons, as illustrated in Figure~\ref{fig:detector_setup}. At the highest nominal beam current, data was also collected with an angle of 90$^{\circ}$ between the direction of the ion beam and the front face of the detector. In the 90$^{\circ}$ orientation the detector is rotated such that the neutron flux is incident on the side of the detector, rather than the front face.

These datasets are listed in Table~\ref{table:data_sets} with their known fluxes as measured using calibrated detectors at PTB. Dataset 4 has a greater flux than dataset 1, despite the beam conditions being the same, due to the greater cross-sectional area the detector presents to the neutron beam in this orientation. The known flux in Dataset 4 is slightly higher than can be attributed to geometric factors alone, as the nominal beam charge for Dataset 4 is 5.6$\%$ greater than in Dataset 1.

\begin{table}[ht]
\caption{Data for the irradiation of the detector in the neutron field with a mean energy of $2.5\1{MeV}$. }\label{table:data_sets}
\begin{center}
\resizebox{\columnwidth}{!}{%
 \begin{tabular}{c c c c r } 
 \hline
Data Set & Current & Orientation & Nominal Charge \newline [$\mu$C] & \;\; Flux [s$^{-1}$] \\
 \hline
 1 & 1.5 $\mu A$ & 0$^{\circ}$ & 2721 & \;16400 $\pm$ 700 \\ 
 2 & 300 $nA$ & 0$^{\circ}$ & 1014 & \;3080 $\pm$ 140\\
 3 & 35 $nA$ & 0$^{\circ}$ & 56.17 & \;340 $\pm$ 15\\
 4 & 1.5 $\mu A$ & 90$^{\circ}$ & 2883 & \;22200 $\pm$ 970\\
\hline
\end{tabular}%
}
\end{center}
\end{table}

The response of EJ-301 to ERs is known to be linear. Data presented in this work is therefore given in terms of the electron recoil equivalent energy keV$_{ee}$. This energy scale is set using the Compton backscatter edge of gammas from $^{60}$Co, $^{137}$Cs and $^{54}$Mn, measured from data collected with the detector in the experimental hall. The background rate of ER events in the experimental hall was measured during an overnight measurement.

A total of 80 million waveforms (amounting to $85\1{GB}$) were collected from the neutron source, background, and calibration gamma-sources, and stored for offline processing.

\section{Discrimination Algorithms}\label{sec:algorithms}

As EJ-301 features different decay constants for NR and ER signals, a variety of methods can be used to discriminate the corresponding waveforms. Five PSD algorithms were implemented in a C++ program to perform offline analysis of the data and compute discrimination parameters for each waveform. These algorithms, described in detail below, are the Charge Comparison Method (CCM), Pulse Gradient Analysis (PGA), Fourier Series Expansion (FSE), Laplace Transform (LAP), and a fit to standard events (SEF). Typical scintillation pulses last for $0.5\1{ns}$ per keV of energy deposited. Each digitized waveform was 525~ns in duration, with the trigger falling between 78 and 94~ns. The first 40~ns were used to calculate a simple baseline average as well as the baseline RMS to indicate the noise level, and the integral of the pulse yields the energy.

\begin{figure}[htbp]
\includegraphics[width=\columnwidth]{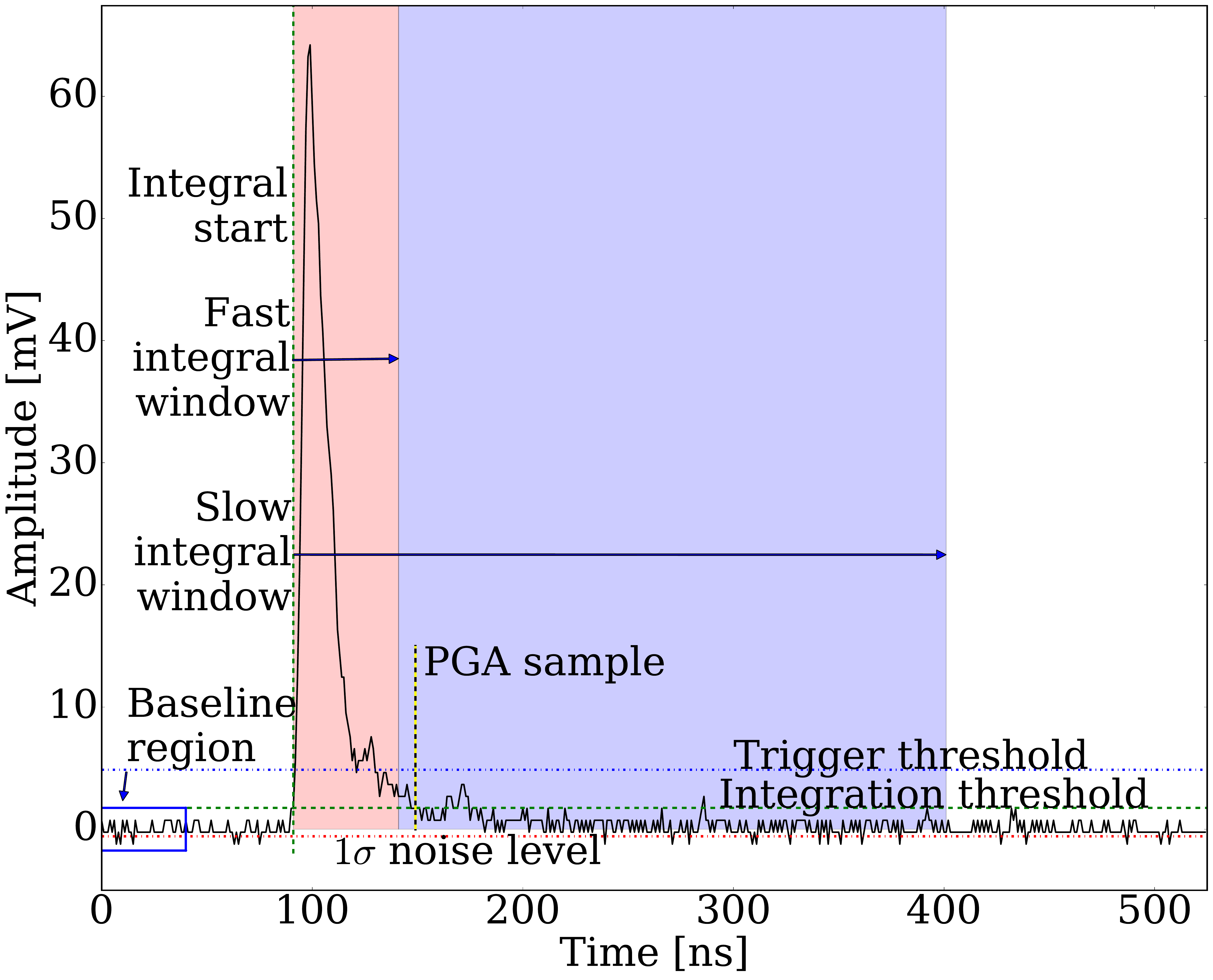}
\caption{A typical gamma event of energy 100 keV. Shown are the trigger and integration thresholds, the ends of the Fast and Slow integral windows, the location of the sample used in the PGA method, the region used for baseline calculations, and the noise level.}
\label{fig:waveform}
\end{figure}

\subsection{Charge Comparison Method}

The Charge Comparison Method (CCM)~\cite{Cester2014,Kaschuck2005,Liao2014,WOLSKI1995,Klein2002,Ranucci1998,Soderstrom2008,RANUCCI1995} predates modern digital computing and was first implemented via passive electronics~\cite{Alexander1961}. In this method, the baseline-subtracted waveform is integrated over two time windows of different lengths, called \textit{slow} and \textit{fast} or \textit{long} and \textit{short}, respectively. The start of these integral windows is the onset of the pulse, which is defined here as the point at which the waveform exceeds 3$\sigma$ of the baseline RMS (as shown in Figure~\ref{fig:waveform}). The lengths of the two windows are generally set to match the decay modes of the detector. As a NR pulse will decay more slowly than an ER pulse, the slow integral value $I_{\n{slow}}$ will be larger for NR waveforms than for ER, while the fast integral values $I_{\n{fast}}$ are typically comparable for both ER and NR waveforms. We have optimized these times according to the traditional Figure of Merit~\cite{ANNAND1987} and found that values of 50~ns for the fast window and 310~ns for the slow window result in optimal discrimination. The discrimination parameter is the ratio of the two integral values,
\[
\n{PSD}_{\n{CCM}} = \frac{I_{\n{slow}}}{I_{\n{fast}}}.
\]

\subsection{Pulse Gradient Analysis}

The Pulse Gradient Analysis method (PGA)~\cite{DMellow2007} compares the relative height of the peak $H_{\n{peak}}$ to that of a sample $H_{\n{sample}}$ a set time after the peak, here 50~ns. This second sample is averaged with the neighboring 10 samples to reduce noise. As ER pulses decay more quickly than NR pulses, the gradient between the peak and this second sample should be larger for the ER than for the NR. We define the discrimination parameter for this method as the ratio between the two amplitudes,
\[
\n{PSD}_{\n{PGA}} = \frac{H_{\n{sample}}}{H_{\n{peak}}}.
\]

\subsection{Fourier Series Expansion}

The third is a method based on a Fourier series expansion of the waveform ($f(t) = \sum_n A_n \exp (i\omega_n t); A_n = \int_0^T \n{d}t\,f(t) \exp (i\omega_n t)$). Using an approach similar to~\cite{Liu2010}, the difference between an even order coefficient $A_{2n}$ (odd order $A_{2n+1}$) and the zeroth coefficient $A_0$ (first coefficient $A_1$) is normalized to the zeroth (first) coefficient and then summed:
\[
\displaystyle
F_{\n{even}} = \sum_{n} \frac{A_{2n}-A_0}{A_0} \n{~and~} F_{\n{odd}} = \sum_{n} \frac{A_{2n+1}-A_1}{A_1}.
\]
The expansion is computed to the 30$^{th}$ order. The discrimination parameter is then defined as the ratio between these two parameters:
\[
\n{PSD}_{\n{FSE}} = \frac{F_{\n{odd}}}{F_{\n{even}}}.
\]

\subsection{Laplace Transformation}

We also evaluate a discrimination algorithm based on the Laplace transform $\mathcal{L}\{f\} = \int_0^{\infty} \n{d}t\,f(t)\exp(-st)$. A 9-point moving average is calculated over the trailing edge of the waveform, and the smoothed pulse is then transformed. The transformed waveform is integrated over two frequency ranges, $0.01\1{GHz}\leq s<0.1\1{GHz}$ and $0.1\1{GHz}\leq s<1\1{GHz}$, denoted `low' and `high', respectively, yielding $L_{\n{low}}$ and $L_{\n{high}}$. The frequency ranges are chosen such that the contributions from the $32\1{ns}$ and $270\1{ns}$ decay modes are maximized in each respective range. The discrimination parameter for this method is defined as $L_{\n{high}}-L_{\n{low}}$. However, since this parameter varies significantly with energy, we choose to re-scale it according to
\[
\n{PSD}_{\n{LAP}} = I- A \log(B + L_{\n{high}} - L_{\n{low}})
\]
in order to aide visualization, with the integral of the pulse given as $I$. Parameters $A$ and $B$ are constants chosen as 1/30 and 2, respectively, simply to linearise the plot.

\subsection{Standard Event Fit (SEF)}

\begin{figure}[htbp]
\includegraphics[width=\columnwidth]{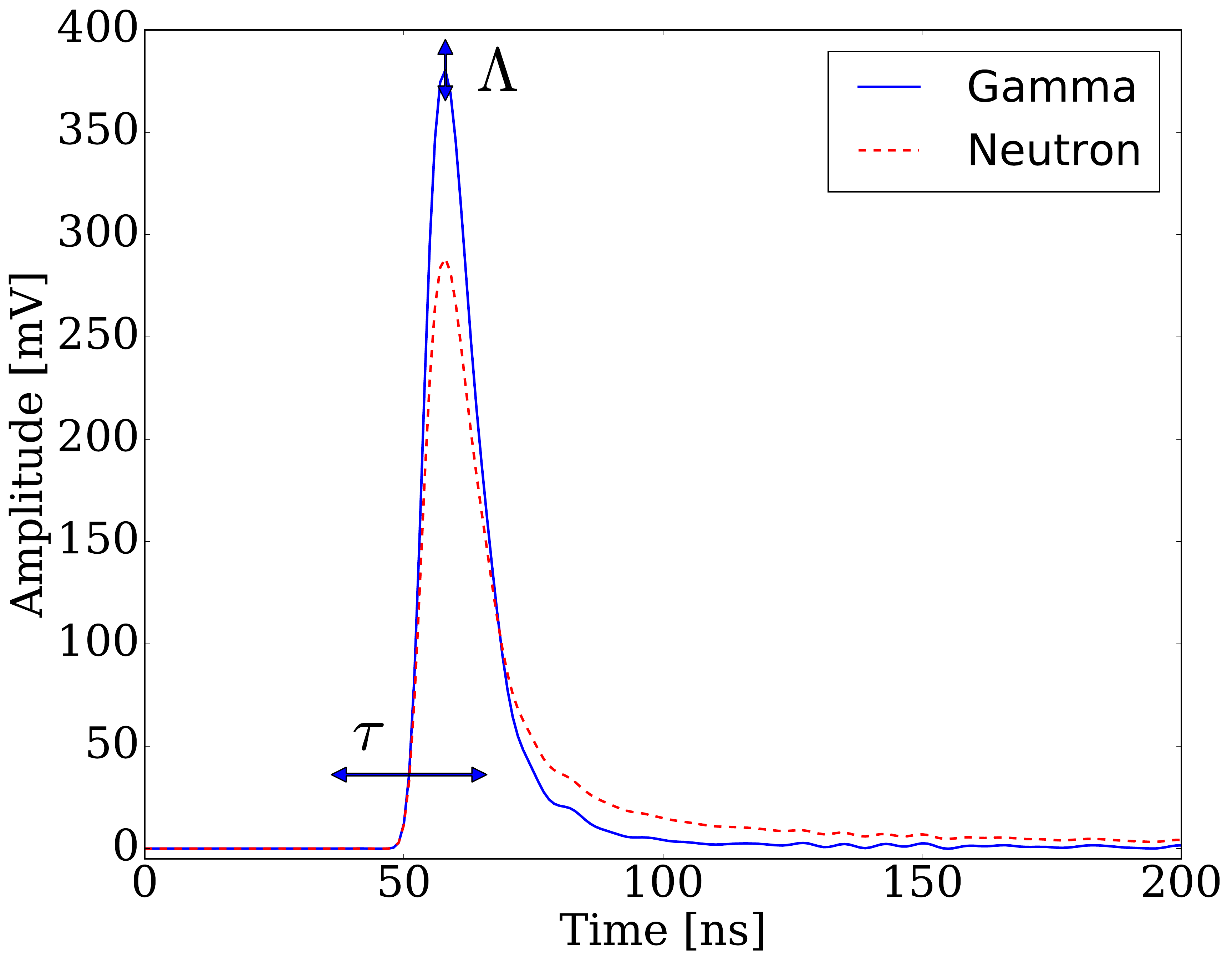}
\caption{Standard nuclear recoil (red dotted line) and electronic recoil (blue solid line) events after baseline subtraction. The free fit parameters are a horizontal shift $\tau$ and a vertical scale factor $\Lambda$}
\label{fig:stdevents}
\end{figure}

Prior to analysis, $10^5$ events identified as NR or ER by CCM are selected in a narrow energy region around the Compton edge of the 662~keV gamma from $^{137}$Cs. Averaging these waveforms together forms the standard NR or ER event. Both these standard events are then fitted to a given waveform~\cite{Guerrero2008,Ambers2011}. While the baseline is fixed to the average of the first 40~ns of the waveform, the free parameters of the fit are a horizontal shift $\tau_{N,E}$ and a scaling factor $\Lambda_{N,E}$, see Figure~\ref{fig:stdevents}. The ER/NR discrimination parameter is then defined as the difference between the chi-squared value $\chi^2_{N,E}$ of each fit, normalized to the vertical scaling fit parameter $\Lambda$:
\[
\n{PSD}_{\n{SEF}} = \frac{\chi^2_N}{\Lambda_N} - \frac{\chi^2_E}{\Lambda_E}.
\]

\section{Analysis}

\begin{figure}[htbp]
\begin{center}
\includegraphics[width=1\columnwidth]{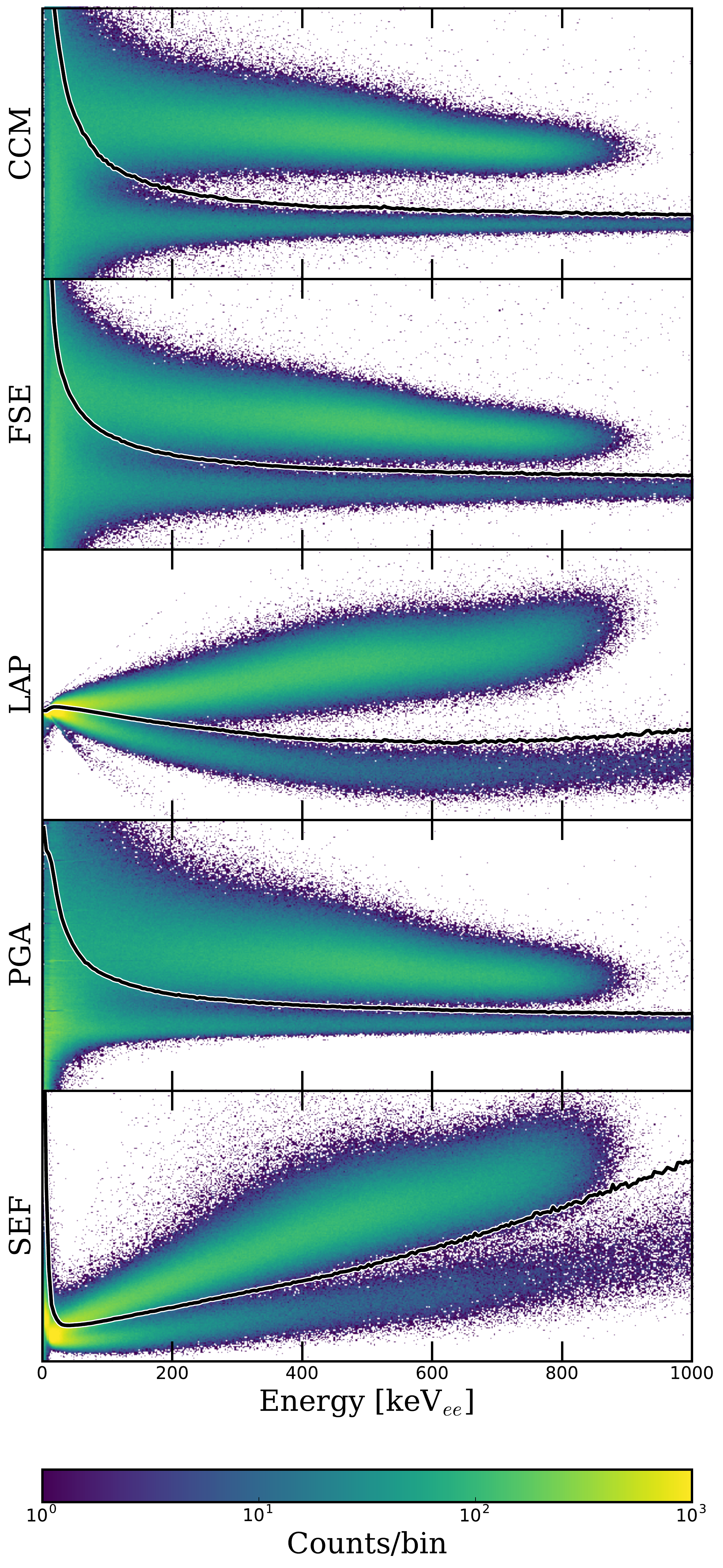}
\caption{Plots of various discrimination parameters versus energy for the CCM~(a), FSE~(b), LAP~(c), PGA~(d), and SEF~(e) algorithms. The 99$\%$ rejection cut of electronic recoils is also shown in each case. The upper populations are the respective nuclear recoil bands.}
\label{fig:discrimination_plot}
\end{center}
\end{figure} 

\subsection{Rejection of Electronic Recoils}\label{sec:backgrounds}
The collected data were each processed using the 5~PSD methods discussed in Section~\ref{sec:algorithms}. Rejection cuts for ER events are defined using the datasets taken with various gamma-sources using histograms of the discrimination parameter versus energy. In each energy bin, we consider the one-sided 95$\%$, 96$\%$, 97$\%$, 98$\%$ and 99$\%$ ER rejection quantiles. Application of the resulting rejection cuts to the overnight background data confirm their performance on neutron datasets. These rejection cuts are robust against changes in detector orientation and data acquisition rate.

The event distribution from neutron data are shown in Figure~\ref{fig:discrimination_plot} for all PSD methods together with the one-sided 99$\%$ ER rejection cut. For ease of comparison, all PSD parameters were defined such that the NR band is the upper event population.

\subsection{Neutron Flux}

Using the rejection levels defined from the ER band, events are tagged as being either an electronic or nuclear recoil. 

\begin{figure}[htbp]
\includegraphics[width=\columnwidth]{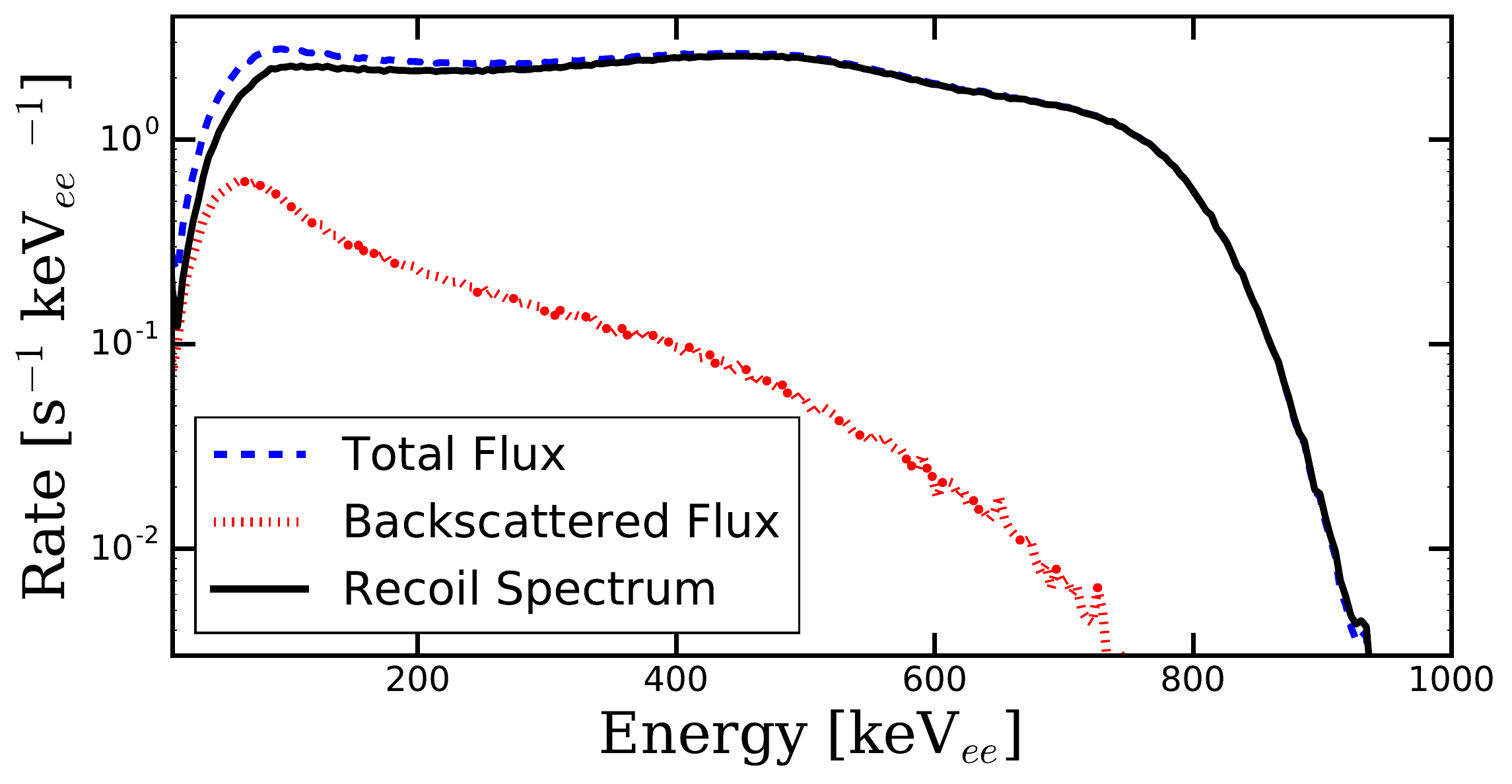}
\caption{Nuclear recoil spectra as measured in the EJ-301 detector using the CCM algorithm. The measured backscatter flux (dotted, red) is subtracted from the total measured flux (dashed, blue) in order to obtain the recoil spectrum of $2.5\1{MeV}$ neutrons (solid, black).}
 \label{fig:rate_plot}
\end{figure}

Figure~\ref{fig:rate_plot} shows the live time-corrected NR energy spectrum, measured from dataset~2, after 99$\%$ rejection of ERs using the CCM~algorithm as an example. The expected double-peaked structure due to neutron double scatter events within the detector volume is evident. 
The recoil spectrum extends to just below $1\1{MeV_{ee}}$, as expected from $2.5\1{MeV}$ neutrons and the non-linear response spectrum of proton recoil energies in EJ-301 detectors~\cite{Verbinski1968, Aksoy1994, Naqvi1993}. Other algorithms show similar spectra with some variations below $200\1{keV_{ee}}$ which will be discussed in section~\ref{sec:acceptance}.

Data collected under the same conditions as dataset~2, but with the shadow cone between the detector and the target is also shown. The observed spectrum is consistent with a range of energies from neutrons scattered off of the air in the experimental hall. The absence of a similar NR spectrum in background data (with no ions on the target), also shown, confirms that these events are related to the beam.

Since we are interested in the efficiency of the EJ-301 detector with regards to detecting $2.5\1{MeV}$ neutrons, we subtract the neutron backscatter rate (Figure \ref{fig:rate_plot}, dotted (red) line) from the total neutron rate (Figure \ref{fig:rate_plot}, dashed (blue) line). This results in the direct neutron flux at the position of the detector (Figure~\ref{fig:rate_plot}, solid (black) line). 

\section{Results}\label{sec:acceptance}

\subsection{Acceptance}

Rather than reducing the performance of different PSD algorithms to a single Figure of Merit value (Eq.~\ref{eq:FOM}), we investigate their energy-dependent behaviour. Specifically, for a given ER discrimination cut, we compare the efficiency of various algorithms using the number of accepted neutrons as a function of energy.

In order to quantify the acceptance of neutrons, a pure band of NRs directly from the beam is required. However, given the reduced performance of all algorithms at low energies, the subsequent overlapping of the NR and ER band, as well as the background from scattered neutrons, no such pure sample is available. We thus invoke a simple statistical algorithm as follows. We use the data taken with the neutron beam incident on the detector
to produce the event distributions in the space of PSD parameter $PSD_i$ versus energy $E$, as shown in Figure~\ref{fig:discrimination_plot}. These distributions are corrected for the livetime of each dataset and the integrated beam current, to obtain the time-normalized event density $\varrho_{\mathrm{sig+bck}}(PSD_i,E)$. Similarly, datasets in which the shadow cone was present result in a time-normalized background density of both ER and NR events $\varrho_{\mathrm{bck}}(PSD_i,E)$. Subtracting these two event densities results in an event density $\varrho_{\mathrm{sig}}(PSD_i,E)=\varrho_{\mathrm{sig+bck}}(PSD_i,E)-\varrho_{\mathrm{bck}}(PSD_i,E)$ that is assumed to be representative of the pure band of NRs directly from the beam. This 2D-histogram of event density is then projected onto the energy axis to obtain a "pure" spectrum of all $2.5\1{MeV}$ neutrons detected by the EJ-301 cell for each algorithm. To check that any mismatch between measurement of ERs with the shadow cone and the background data does not bias the pure neutron flux, we subtract the background rate from the ER rate measured with the shadow cone. The resulting spectrum is found to contribute at most 0.1$\%$ uncertainty to the assumed pure neutron flux directly from the beam.

\begin{figure}[ht]
\centering
\includegraphics[width = \columnwidth]{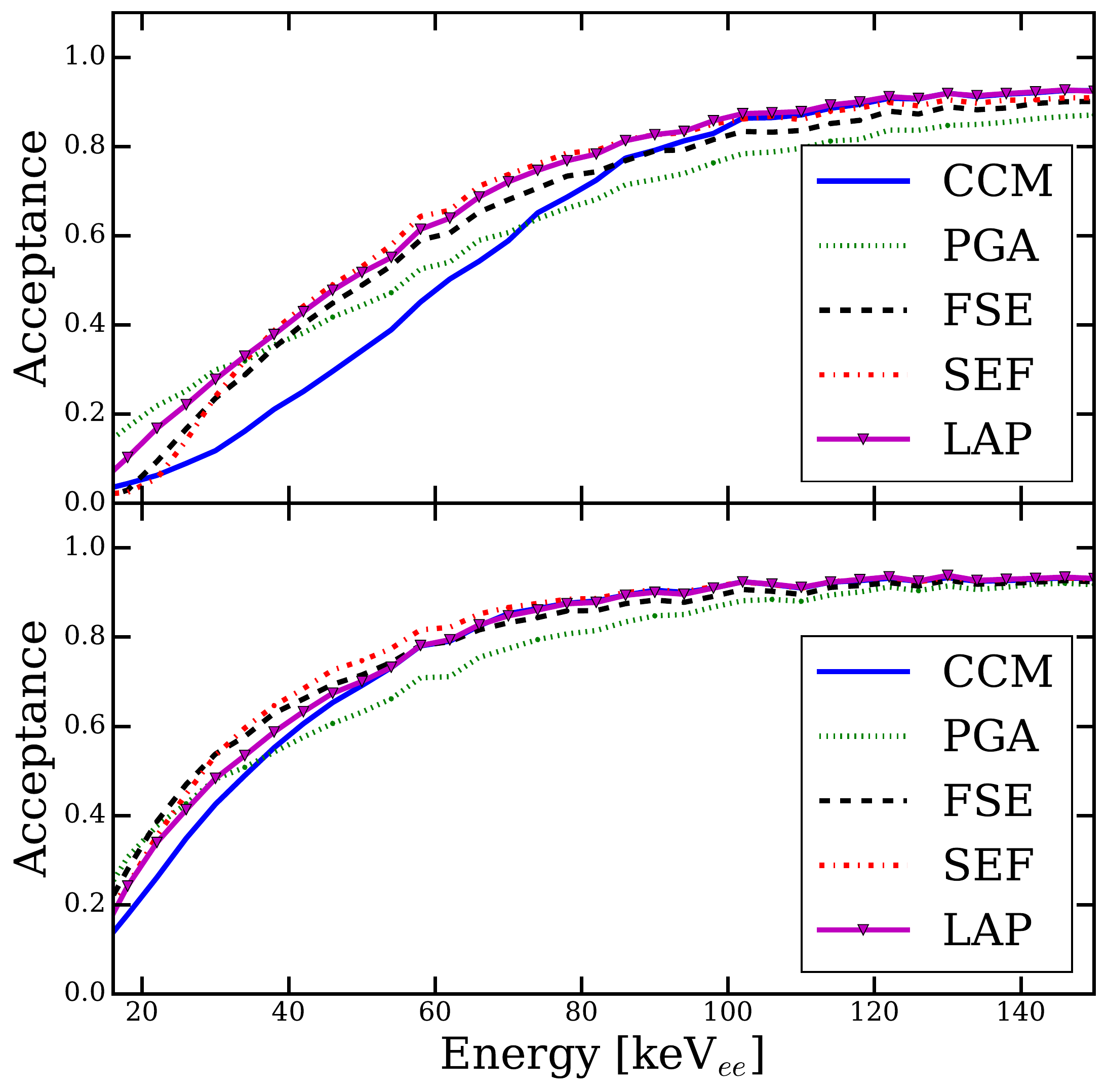}
\caption{Energy dependence of the fraction of nuclear recoil events that are accepted by each algorithm after 99$\%$ (top) and 95$\%$ (bottom) rejection of electronic recoils.}
\label{fig:acceptance}
\end{figure}

The acceptance of neutrons is determined by dividing the measured NR spectrum for a given dataset and algorithm (as shown in Figure~\ref{fig:rate_plot}) by this "pure" NR spectrum. The resulting energy-dependent acceptance of neutrons at a constant rejection level of ERs is shown in Figure~\ref{fig:acceptance}.

In the low-energy region shown in Figure~\ref{fig:acceptance}, the acceptance of the SEF~algorithm performs best above $30\1{keV_{ee}}$ at 95$\%$ rejection of ERs. The acceptances of the LAP and FSE algorithms match each other at this rejection level. As the rejection level of ERs is increased to 99$\%$, the acceptance of the LAP algorithm at higher energies is marginally but consistently higher than that of the SEF algorithm. Of note here is that the performance of the traditional CCM algorithm decreases as the rejection criteria for ER events becomes more stringent. Specifically, at 95$\%$ rejection of ERs, its acceptance is consistent with that of the SEF, FSE and LAP algorithms, and better than the PGA algorithm. However, at 99$\%$ rejection of ERs, the CCM algorithm has the lowest acceptance below $80\1{keV_{ee}}$.

\subsection{Purity of the Nuclear Recoil Spectrum}

\begin{figure}[htbp]
\centering
\includegraphics[width = \columnwidth]{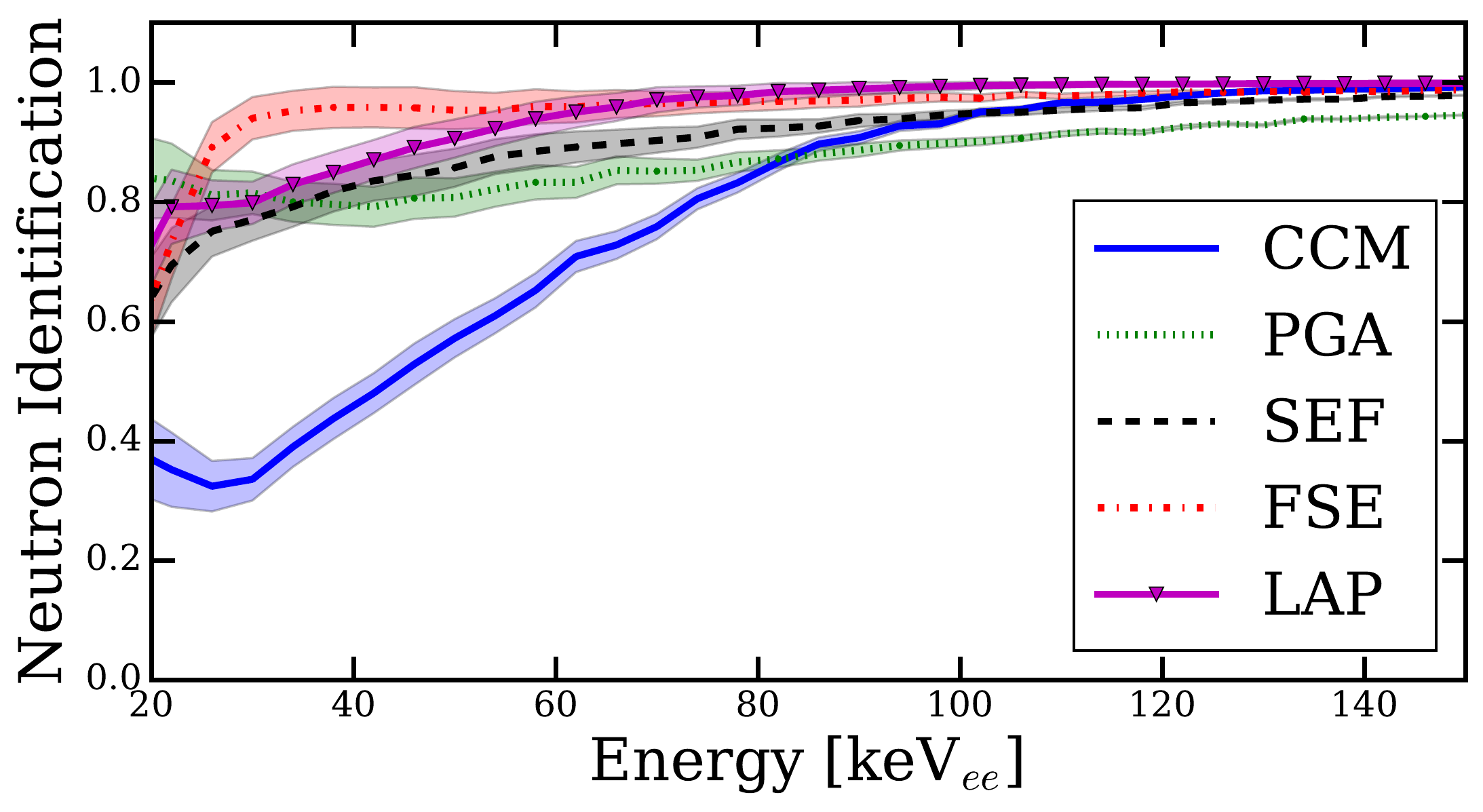}
\caption{The fraction of NRs identified by a given algorithm from a reference sample. The reference sample consists of events identified as NRs by all algorithms except the algorithm under consideration. The shaded region indicates the systematic uncertainty.}
\label{fig:n-1_comparison}
\end{figure}

At low energy ($<100\1{keV_{ee}}$), there is considerable overlap between the NR and ER bands. Therefore, we not only want to consider the ability of an algorithm to accept neutrons, but also the purity of the resulting NR spectrum. To this end, we construct a reference sample for each algorithm $i$ that contains events that are NRs with high confidence, using those events that are identified by all other algorithms $j\neq i$ as a NR. As a function of energy, we then calculate how many of the events in the reference sample were also tagged as a NR by the algorithm under investigation. The fraction of events in the reference sample that an algorithm has tagged as a NR is shown in Figure~\ref{fig:n-1_comparison}.

The inability of the traditional CCM algorithm to identify NR events in a reference sample below $100\1{keV_{ee}}$ is striking. In contrast, the SEF algorithm performs best at low energies and correctly identifies more than 95$\%$ of NR events down to energies below $50\1{keV_{ee}}$. For the SEF algorithm, this fraction is flat down to energies well below the point at which the acceptance of neutrons for all algorithms has fallen below 50$\%$ (compare Figure~\ref{fig:acceptance}). Intriguingly, while the LAP algorithm does provide lower efficiency at low energies, it provides a slightly higher level of confidence that an event is truly a neutron above $\sim75\1{keV_{ee}}$. In contrast to~\cite{DMellow2007}, we find that the performance of the PGA algorithm is inferior to those of all other algorithms above $80\1{keV_{ee}}$. This observation can be attributed to the fact that the PGA algorithm is based on only two samples of the recorded waveform, thus being highly susceptible to electronic noise.

The performance of algorithms which perform poorly at low energies, such as the CCM algorithm, can bias the selection of the reference neutron sample. To estimate the systematic uncertainty in the fraction of identified NRs, we neglect one algorithm, and repeat the procedure above, i.e. we construct reference samples from three of the four remaining samples, and examine which fraction of NRs are identified by the fourth method. In this way we obtain four neutron identification curves for each method, from which we determine the standard deviation. This is taken as a measure of the systematic uncertainty of each method and is shown as the shaded regions around each curve in Figure~\ref{fig:n-1_comparison}.

\subsection{Detector Efficiency}

\begin{figure}[h!]
\centering
\includegraphics[width = \columnwidth]{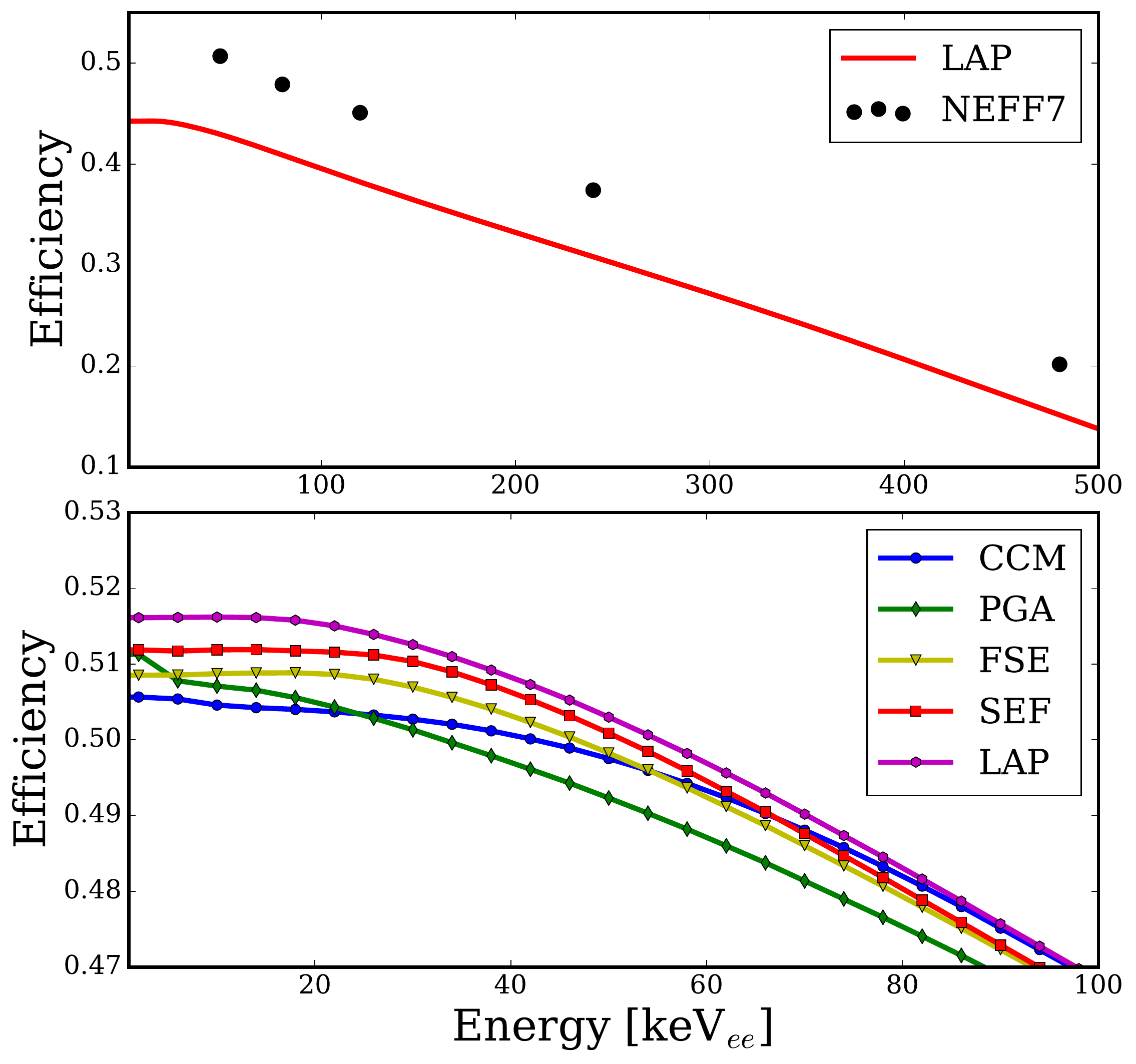}
\caption{(Top) Comparison between the threshold-dependent efficiency of the SEF algorithm and the prediction from the NEFF7 code in EJ-301. The detector is orientated at 90$^{\circ}$. (Bottom) The efficiency of all algorithm as a function of threshold. Note the different energy ranges.}
\label{fig:efficiencies}
\end{figure}

Table~\ref{table:data_sets} lists the total expected neutron flux of $2.5\1{MeV}$ neutrons through the detector for each dataset. Based on the determined absolute neutron rates from the PTB measurements, we reconstruct the detection efficiency of the detectors as function of energy threshold. The results at 99$\%$ rejection of ERs are shown in Figure~\ref{fig:efficiencies}. In addition, the expected efficiency from a MC simulation using the NEFF7 code~\cite{GH1982} is shown in Figure~\ref{fig:efficiencies} for the arrangement in which the detector orientation was 90$^{\circ}$ at selected energy thresholds. We find that the efficiency measured in the detector is inconsistent with both the overall efficiency described by the NEFF7 code, as well as the functional dependence of the efficiency on the threshold. Independent of the PSD algorithm, the measured efficiency in the EJ-301 detector is lower and less threshold-dependent than that described by the NEFF7 code. We thus conclude that the NEFF7 code can currently not be used to obtain accurate detection efficiency information for EJ-301 liquid scintillator cells.

The PGA algorithm has the lowest efficiency above $30\1{keV_{ee}}$, consistent with it having a poorer ability to accept neutrons at a given rejection level of ERs. The traditional CCM algorithm has poor efficiency below $60\1{keV_{ee}}$ when compared to the three newer algorithms FSE, SEF and LAP. The LAP algorithm has the highest efficiency at all energy thresholds. This observation can be ascribed to its higher acceptance of NRs at higher energies. The LAP algorithm is better at discriminating between the NR and ER band at higher energies, as the ER band of the SEF algorithm spreads as the energy increases (compare Figure~\ref{fig:discrimination_plot}), while the ER band of the LAP algorithm has a more consistent width.

\subsection{Processing Speed Benchmarks}

The processing speed of each algorithm was measured on a dataset of approximately $500\,000$ events (about $550\1{MB}$), both with and without the C++ compiler optimizations available in the GNU compiler collection version 4.8.3, using a standard desktop computer. The results are show in Table~\ref{tab:benchmarks}. From this we see that the increased low-energy performance of the SEF algorithm comes at a very high computational cost.

\begin{table}[htbp]
\begin{center}
\caption{Processing rates of the different algorithms, given in units of waveforms/sec, using a standard desktop computer.}\label{tab:benchmarks}
\begin{tabular}{c p{26mm}p{26mm}}
 \hline
Algorithm & Rate without \newline optimizations & Rate with \newline optimizations \\ 
\midrule
PGA & $1.6\times10^6$ & $4.8\times10^6$ \\
CCM & $168\,000$ & $1.6\times10^6$ \\
FSE & $4\,300$ & $84\,000$ \\
LAP & $1\,100$ & $16\,000$ \\
SEF & 110 & 230 \\ \bottomrule
\end{tabular}
\end{center}
\end{table}

\section{Conclusions}

We have compared the performance of five different pulse shape discrimination algorithms using a commercial liquid scintillator cell. The studied algorithms include the avant-garde algorithms Standard Event Fit SEF, Fourier-Series Expansion FSE, and Laplace Transform LAP in addition to the traditional Charge Comparison Method CCM and Pulse-Gradient Analysis PGA. The energy-dependent behaviour of all five algorithms was discussed as a better means of describing PSD algorithms than the Figure of Merit previously used in the literature.

Specifically, we considered the ability of each algorithm to accept NRs as function of recoil energy, as well as the purity of the resulting NR sample. We find that at 99$\%$ rejection of ER background events and above $80\1{keV_{ee}}$, the PGA algorithm accepts the fewest number of NRs and is the least likely to identify a NR event from a reference sample. The CCM algorithm performs better than the PGA method, but a marked deterioration in the performance of the CCM algorithm is observed as the rejection level of ER events becomes more stringent.

Both the SEF algorithm and the LAP algorithm display improved performance compared to the traditional methods considered. The SEF algorithm is more likely to accept NRs from a reference sample at low energies, and it provides a higher acceptance of NR events below $80\1{keV_{ee}}$. The LAP method however provides a slightly higher efficiency overall for the detection of NRs in EJ-301 due to a better-resolved ER band at higher energies. 

\section{Acknowledgements}

This work was supported by grant \#PHYS-1412965 from the National Science Foundation (NSF) and carried out under a cooperation agreement between Purdue University and PTB. JP~is supported by scholarship \#SFH13071722071 from the National Research Foundation (NRF).

\end{document}